\documentclass[onecolumn]{aa2}
\usepackage{graphicx}

\begin{document}
\title{\vspace{-1.4cm}\\
An almost head-on collision as the origin of two off-centre rings in the Andromeda galaxy\vspace{1.2cm}}
\author{
D.L. Block\inst{1}, F. Bournaud\inst{2,3}, F. Combes\inst{2}, R.
Groess\inst{1}, P. Barmby\inst{4}, M.L.N. Ashby\inst{4}, G.G. Fazio\inst{4}, M.A.~Pahre\inst{4},
S.P. Willner\inst{4}} \offprints{D.L. Block \email{block@cam.wits.ac.za}}

\institute{ Anglo American Cosmic Dust Laboratory, School of
Computational and Applied Mathematics, University of the
Witwatersrand, Private Bag 3, WITS 2050, South Africa \and
Observatoire de Paris, LERMA, 61 Av. de l'Observatoire, F-75014,
Paris, France \and Present Adress: DSM/DAPNIA/Service
d'Astrophysique, CEA/Saclay, 91191 Gif-sur-Yvette Cedex, France \and
Harvard-Smithsonian Center for Astrophysics, 60 Garden Street,
Cambridge, Massachusetts
 }

\date{Received 24 May 2006 ; Accepted 18 August 2006 ; Published in Nature, October 19th issue.}

\abstract{The unusual morphology of the Andromeda Spiral (Messier
31, the closest spiral galaxy to the Milky Way) has long been an
enigma. Although regarded for decades as showing little evidence of
a violent history, M~31 has a well-known outer ring$^{1–-7}$ of star
formation at a radius of 10~kpc whose center is offset from the
galaxy nucleus. In addition, the outer galaxy disk is warped as seen
at both optical$^8$ and radio$^9$ wavelengths. The halo contains
numerous loops and ripples.  Here we report the discovery, based on
analysis of previously-obtained data$^{10}$, of a second, inner dust
ring with projected dimensions 1.5 by 1~kpc and offset by $\sim$
0.5~kpc from the center of the galaxy. The two rings appear to
be density waves propagating in the disk. Numerical simulations
offer a completely new interpretation for the morphology of  M~31:
both rings result from a companion galaxy plunging head-on through
the center of the disk of M~31.  The most likely interloper is M~32.
Head-on collisions between galaxies are rare, but it appears
nonetheless that one took place 210 million years ago in our Local
Group of galaxies.}
\authorrunning{Block et al.}
\titlerunning{Two collisional rings in M~31}
\maketitle

Newly-acquired images$^{10}$ of M~31 secured by the Infrared Array
Camera$^{11}$ (IRAC) onboard the Spitzer Space Telescope span the
wavelength regime of 3.6 to 8.0 microns. These images offer unique
probes of the morphologies of the stellar distribution and
interstellar medium with no interference from extinction. Figure 1
shows the emission map of the interstellar medium at 8 microns,
generated by subtracting a scaled 3.6 micron image (dominated by
starlight) from the 8 micron image. The subtraction removes the
contribution from stellar photospheres and leaves only the emission
from dust grains$^{10}$, which trace the interstellar medium of M~31.
What is most striking in Figure 1 (and in the enlarged inset) is the
presence of a complete – though asymmetric – inner ring of dust 6.9
by 4.4 arcmin in extent, translating to linear dimensions of about
1.5 by 1 kpc (assuming a distance$^{12}$ of 780 kpc). The inner ring
lies between the two well known Baade spiral dust arms$^{13}$, both
of which are clearly seen in emission. The inner ring is elongated
in a direction close to the minor axis and belongs to the central
gas disk, which appears to be more face-on$^{14}$. It is therefore
not possible to know the inner ring's precise ellipticity, but it is
unlikely to be circular. IRAC imaging thus reveals two rings. The
outer ring is offset by approximately 10 percent of its radius,
while the inner ring is offset by about 40 percent or ~0.5 kpc. The
inner elliptical ring has been alluded to in earlier
studies$^{15,16}$, but all investigators have hitherto believed it
to be a mini-spiral, related to a bar. Published Spitzer 24 micron
images5 of M~31 show centrally-concentrated dust emission; the ring
morphology is therefore disguised at these longer wavelengths. The
IRAC images beautifully show the inner ring at high spatial
resolution and furthermore confirm that this feature is a complete
and continuous ring, even though offset and asymmetrical. There are
two known scenarios whereby disk systems form rings: by head-on
galaxy collisions or by rotating bars. Head-on collisions differ
from common tidal interactions between galaxies because the
pericentric distance of the orbit of the companion is small and its
orbit is almost perpendicular to the target disk. Such collisions
produce expanding ring-shaped waves$^{17}$. Rotating bars in spiral
disks produce inner, outer, and sometimes nuclear rings in barred
spiral galaxies. The rings occur at orbital resonances and arise
from the galaxy's internal dynamics (unlike collisional rings). Many
examples of bar-induced rings are known$^{18}$.
\medskip

\begin{figure}
\centering
\resizebox{18cm}{!}{\includegraphics{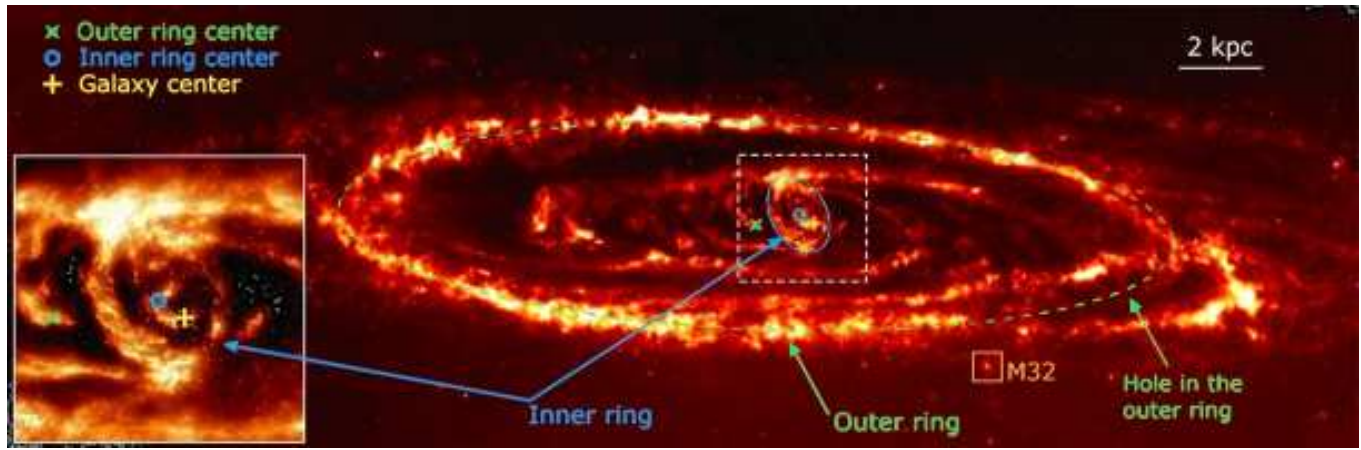}}
\caption{The Andromeda Galaxy M~31 observed with the Infrared Array
Camera$^{11}$ on board the Spitzer Space Telescope at a wavelength of 8~microns.
A scaled version of the 3.6 micron image was subtracted from the 8~microns image
to remove light from stellar photospheres. The remaining emission therefore
traces the emission of warm dust grains and macromolecules in the interstellar
medium. M~31 clearly possesses two rings. Apart from the famous outer dust ring
seen at a radius of ~10 kpc, this map reveals a second 1.5 by 1 kpc inner dust
ring offset by approximately 0.5 kpc from the galaxy nucleus. Both rings are
interpreted to be density waves induced by an almost head-on collision. The most
likely candidate is the dwarf companion galaxy M~32, which appears faint in this
image because it has little dust. The starlight of M~32 is however prominent in
the separate 3.6 and 8 micron images.}
\end{figure}

Infrared imaging of spiral galaxies in the Local Universe confirms
the ubiquity of the bar phenomenon; about 80\% of spiral galaxies
are barred$^{19}$, even if many are classified as unbarred at
visible wavelengths. M~31 is no exception; although Hubble classified
it as normal type Sb, it is likely to possess a stellar bar. The
presence of a bar in M~31 is suggested by its boxy bulge, which is
conspicuous in infrared imaging$^{5,20}$ and is aligned almost
parallel to the galaxy's major axis. The boxy bulge might be an
inflated bar, or it may hide a thinner parallel bar (at about 10
degrees from the major axis). There is, however, no bar parallel to
the minor axis, along which the major axis of the inner dust ring
lies. Had there been a stellar bar along the minor axis, it would
surely have been detected in the IRAC 3.6 micron image10, which is
much deeper than the 2.2 micron 2MASS images and less affected by
dust attenuation. The double-ring system of M~31 therefore appears
unrelated to any possible central bar, because the elliptical inner
ring does not have the typical orientation of most bar-induced gas
rings (generally aligned along the bar's major axis). Moreover, the
off-center ring strongly supports the collision interpretation.
Resonant rings generated from rotating bars are rarely as off-center
as the rings of M~31, and the amount to which such rings show an
off-center increases with radius. In M~31, however, the inner ring
is off-center by 40\% of its radius whereas the outer ring is
off-center by only 10\%.
\medskip

The relative brightness of the inner and outer rings further
supports a collision origin. When a bar induces a strong pair of
rings, the most prominent one is generally the inner, located just
beyond the extremity of the bar at the so-called 4:1 resonance. In
M~31, the outer ring is much brighter than the inner one. This makes
the interpretation of bar-induced resonance rings highly unlikely
except under the unrealistic hypothesis of a very slowly-rotating
bar extending up to the outer ring (contrary to observations). The
dual-ring morphology of M~31 strongly indicates expanding
ringeddensity waves triggered by a head-on galaxy collision with a
companion.
\medskip

The dwarf companion galaxy M~32 is very likely to be the impactor. To
investigate whether a collision of M~32 with the disk of M~31 could
induce the two-ringed density waves identified in the IRAC images,
N-body simulations were performed$^{21}$. The simulations include
stars, gas, and dark matter in M~31 and M~32 with one million
particles and a spatial resolution of 350 parsecs. Our simulations
differ from all previous ones in that the initial orbit of M~32 is
close to the polar axis of M~31, and the mass ratio is larger.
Previous simulations$^{5}$ have assumed that the impact occurred
several kiloparsecs away from the center of M~31 and not along the
rotation axis (nearly head-on). Our new simulations start with an
axisymmetric disk, proceed ~1 Gyr while a bar and spiral arms form,
then introduce a collision with M~32. We not only take into account
the mass stripped during the collision but also take cognizance of
the dark matter associated with M~32 itself. We assume an initial
mass ratio for M~32 of 1/10 that of M~31 (including dark matter).
Current-epoch mass estimates of M~32 are lower, but much mass is
likely to have been stripped$^{22}$ during the collision. The final
mass of M~32 in our model is 1/23 that of M~31, compatible with
present-day mass estimates.
\medskip

\begin{figure}
\centering
\resizebox{10cm}{!}{\includegraphics{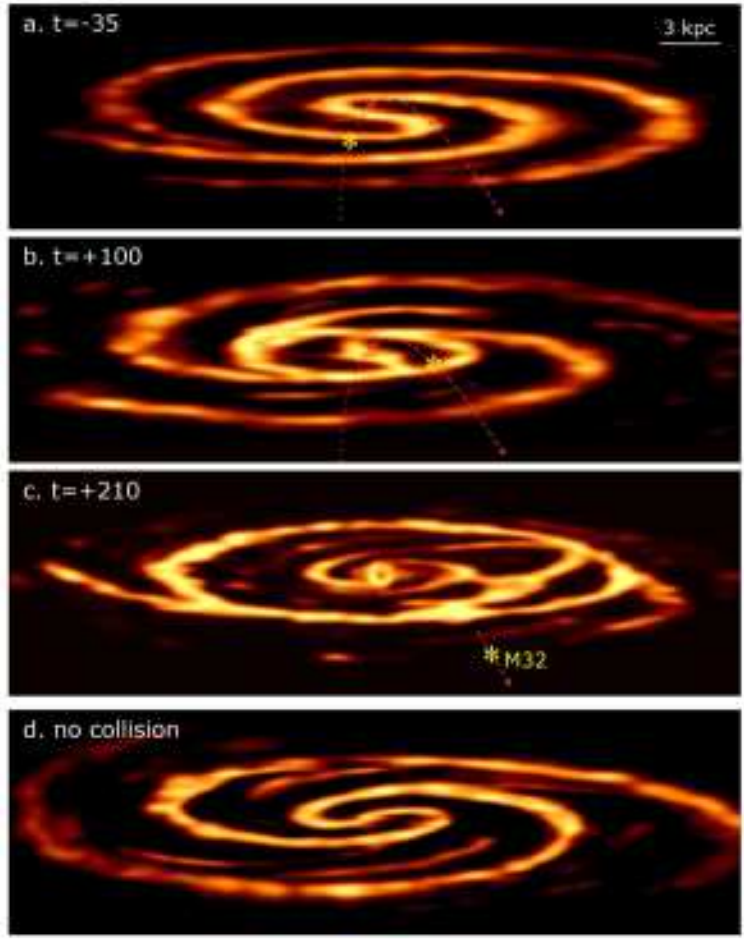}}
\caption{Gas morphology produced by simulations$^{21}$ of a head-on encounter
between M~31 and M~32. These N-body models include the gravitational dynamics
of stars, interstellar matter, and dark matter in both M~31 and M~32. A
''sticky-particle'' scheme accounts for the dissipative nature of the
interstellar medium, and star formation is also included21.  The dashed
red line demarcates the orbit of M~32; the locale of impact lies very close
to the polar axis of M~31. Snapshots a, b, and c occur at t= 35 million
years (Myr) prior to collision and at 100 and 210 Myr post impact; the latter
snapshot also shows the position of M~32 as we see it today. All snapshots
are with the disk viewed at an inclination angle of 77 degrees for direct
comparison with Figure 1. Snapshot c shows the central region of M~31 warped
by a tilt angle of 30 degrees with respect to the main disk plane in accord
with observations$^{14}$.  The two ring-like waves (both offset from the galaxy
center) are seen, as is the hole in the outer ring. Snapshot d shows the gas
morphology at t=210 Myr starting from the same initial disk conditions but
modeled without a collision; no density-wave rings are generated. The initial
mass ratio for the companion M~32 is 1/10th that of M~31 (or 1/13th excluding
dark matter).  The companion is assumed to impact the M~31 disk at a velocity
of 265 km/s with an impact parameter of 4 kpc. It would now be located at a
distance of 35 kpc and at a galactic latitude of about 45 degrees, which is
fully compatible with the present position$^{30}$ of M~32.}
\end{figure}

The simulation of a head-on collision (Figure 2) beautifully
reproduces the observed global morphologies of M~31 reported in this
Letter. The model produces two striking density wave rings offset
from the galaxy center. The outer density wave propagates through
the M~31 disk, resulting in a perturbed interlacing of spiral
features with an outer ring. This produces the conspicuous hole
observed in the outer 10 kpc ring of M~31 (compare Figures 1 and 2).
Our model reproduces this very striking (but transient) feature and
adds compelling evidence that the two rings have been induced by a
collision 210 million years ago close to the rotation axis of M~31.
The non-zero (but small) impact parameter produces the off-center of
the inner ring, and the inclination of the orbit with respect to the
rotation axis produces the ring elongation. In the simulations, the
head-on collision does not destroy the pre-existing boxy bulge in
the barred target disk but merely weakens it.
\medskip

Collisional ring galaxies are expected to show local radial and
tangential perturbed motions in the vicinity of the rings, which
will be observed as ''streaming motions''.  When the intruder is
one-fourth or less as massive as the disk galaxy, the ring behaves
like a wave which propagates through the disk$^{23}$, as opposed to
the large-scale bulk motion of material that occurs when the mass of
the intruder is larger. Our model predicts radial velocities in the
gas of 10 km/s at the outer (10 kpc) ring. Unwin$^{24}$ finds local
streaming velocities in neutral hydrogen gas of 30 km/s at that
radius; our predicted radial velocities constitute 30 percent of the
total observed streaming velocity. The inner ring is predicted to
show lesser radial velocities,   which are more difficult to probe
because of the vertical tilt of the central gas disk. There is an
interesting  kinematic signature of a ring induced by collision:
the resulting positive radial velocity in the ring is predicted to
be associated with a depression in the rotational velocity towards
the outer parts of the ring (with a corresponding increase in the
inner parts of the ring). This is indeed seen in most isovelocity
contours in neutral hydrogen$^{24}$. The predicted tangential
streaming motions are actually opposite to the sense expected for
spiral arm density waves$^{25}$. Supplementary information is linked
to the online version of the paper at www.nature.com/nature (See Appendix in this preprint).
\medskip

The most striking analog of the disk morphology observed in M~31 is
the Cartwheel Galaxy. The Cartwheel is the archetype of a
double-ringed morphology produced by collision. Two distinct density
wave rings are observed: a conspicuous outer ring, associated with
massive star formation, and an elliptical inner ring, offset from
the center of the galaxy. In between the two rings lie several
spiral arms, termed "spokes", which have developed in a trailing
pattern. Several models have been computed of the Cartwheel Galaxy,
simulating the almost head-on collision with one of its
companions$^{26,27}$.
\medskip

M~31 contrasts with the Cartwheel Galaxy in that we do not see the
inner ring in starlight but only in gas and dust. In other words,
the inner ring in the Cartwheel Galaxy shows a much greater degree
of contrast (compared to the parent disk) than does the inner ring
in M~31. The explanation is that the impactor in the Cartwheel Galaxy
must have been more massive than M~32, resulting in two density-wave
rings of stars. Our simulations match the lower contrast seen in
M~31: the primary ring wave propagates outward in the disk, being
created by a crowding of particle trajectories and triggered massive
star formation at the peak of the wave. The second elongated
density-wave ring begins to propagate in the central regions of the
galaxy, most likely in a tilted central disk. At the present epoch,
the gas morphology of the model agrees with the double-ring
structure of the M~31 interstellar medium as revealed by
Spitzer/IRAC.
\medskip

The peculiar morphology of M~31 has been mysterious for many years,
but the discovery of the offset inner ring may be the clue needed to
offer an explanation: a recent head-on collision can produce both
the inner ring and the previously-known outer one. The rings are
unlikely to have been created by bar resonances because of the
rings' relative brightness and their offsets from the galaxy center
and also because they show no relationship to the spiral structure;
in particular, no minor axis bar is observed. While head-on
collisions between galaxies may have been common in the early
Universe$^{28,29}$, only a handful are known nearby. The discovery
of one in our near neighbor M~31 affords the unique opportunity of
studying such a collision at unprecedented spatial resolution.
\bigskip

{\bf Appendix -- Supplementary information: The velocity field and streaming motions induced by the head-on collision}
\smallskip

Ring waves induced in a target galaxy disk by a head-on collision produce
specific velocity perturbations, which may provide a signature of the event.
Models show how complex these perturbations are when all three dimensions are
taken into account$^{17}$. In addition to the perturbation in the rotational (tangential)
and radial velocities, there are characteristic velocities perpendicular to the disk
plane, which becomes warped and corrugated. Fortunately, due to the high inclination
of M~31, the velocity perturbations perpendicular to the plane will have little
effect on the Doppler velocities observed from our direction.
\medskip

For small perturbations, such as the one proposed here (where the companion mass
is less than 1/10 of the target mass), it is possible to model the perturbation
analytically with first order epicyclic theory: after receiving a velocity impulse
due to the passage of the companion through the target galaxy, the particles in
the target disk execute small oscillations around a guiding centre in a circular
orbit. Most of the effect at the beginning of the perturbation can be modelled by
the resulting kinematic wave, the self-gravity of the perturbation only becoming
important at later epochs. The epicyclic approximation considers the induced
oscillations to be harmonic, with a small variation $\delta r$ in galactocentric radius:
\begin{equation}r = r0 + \delta r \cos \kappa t \end{equation}
and in azimuth
\begin{equation}\Theta = \Omega t -2 \Omega/\kappa \delta r/r \sin \kappa t \end{equation}
The galactocentric radial velocity is then
\begin{equation}V_r = - \kappa r \sin \kappa t \end{equation}
and the perturbation in tangential velocity is
\begin{equation}V_t= -2 \Omega \delta r \cos \kappa t \end{equation}
($\Omega$ and $\kappa$ are the rotation and epicyclic angular
frequencies, respectively).
\medskip

At a particular epoch, an induced ring wave appears where particles expanding
outwards in their epicycle oscillation encounter particles in other orbits flowing
inward. The sense of the dominating radial velocity of most particles in the ring
depends on the precise potential and variation of the epicyclic frequency $\kappa$ with
radius. For a flat rotation curve (as appropriate to M~31 at a radius of 10 kpc),
the predicted radial velocity is positive. The magnitude of this radial velocity
can be estimated from knowledge that $\kappa$ is $\simeq 1.4 \Omega$ for a flat rotation curve. With
a companion whose mass is 10\% of the target disk mass, the maximum amplitude $\delta r/r$
is 10\% in the centre, where the companion impacts the disk. In the outer parts, where
the outer ring is currently located, the amplitude should be less, of order 5\%. The
maximum radial velocity perturbation is then $0.05 \times 1.4 \Omega r$, or equivalently 0.07
$V_{rot}$, where $V_{rot}$ is the rotational velocity. If $V_{rot}$ = 200 km/s, we expect a {\it maximum}
radial velocity of 14 km/s. The observed velocity will be less because not all
particles in the ring are in outward motion, and not all those are at their maximum
velocity. Thus the radial kinematical effect one can realistically expect is 0.7
times the maximum or 10 km/s. The numerical model confirms this analytic estimate,
as shown in Figure~3.
\medskip

\begin{figure}
\centering
\resizebox{10cm}{!}{\includegraphics[angle=270]{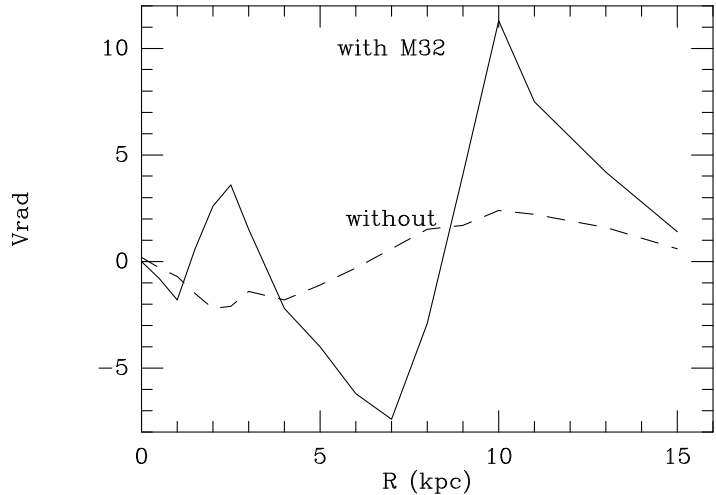}}
\caption{Predicted ring-induced galactocentric radial velocities
(averaged azimuthally) in our numerical model of M~31. The solid
line shows radial velocities in our model after the collision with
M~32. The dashed line shows the unperturbed disk of M~31 with only
its underlying spiral structure.}
\end{figure}

\begin{figure}
\centering
\resizebox{10cm}{!}{\includegraphics{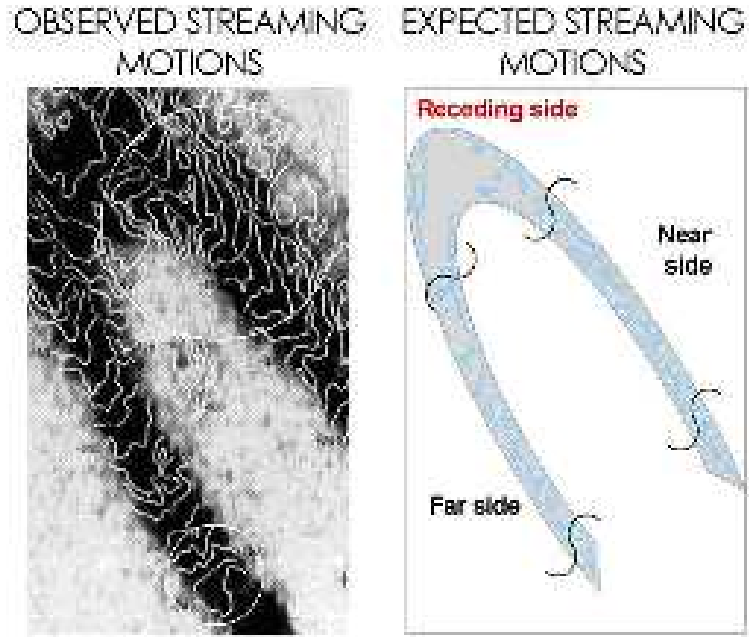}}
\caption{Isovelocity curves observed in neutral hydrogen gas$^{24}$ compared with predictions
from our collisional model. The contour interval is 10 km/s. The northern part of M~31 is
the receding part (redshifted). The near side is NW and the far side SE. The expected
wiggles are indicated for the regions near the minor axis (radial velocity outwards for
the outer part of the ring) and for the regions near the major axis (depression of
tangential velocity in the outer part of the ring). The ''S'' curves schematically depict
expected Doppler velocities; only the portion within the ring area (shaded blue) can be
observed. The regions where the streaming motions support the collisional model are circled in white.}
\end{figure}

\begin{figure}
\centering
\resizebox{10cm}{!}{\includegraphics{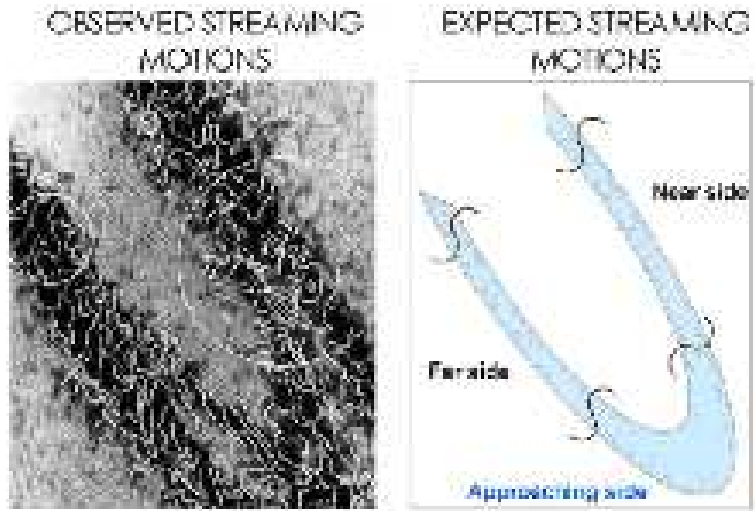}}
\caption{Same as for supplementary Fig.~4, but for the southern sector of M~31, which is approaching the observer (blue-shifted).
}
\end{figure}

As noted in the previous paragraph, the epicyclic approximation predicts larger
perturbations in the tangential velocities for a flat rotation curve by a factor
of 1.4. In the numerical model, tangential streaming motions are indeed seen ranging
between 15 and 20 km/s at the 10 kpc ring radius. The sense of the dominating
velocities in the ring wave depends on the potential, and a gradient of velocity
is predicted at the ring crossing. When the companion crosses the target disk,
particles are given an inward pull, and the conservation of angular momentum requires
that tangential velocities increase with respect to the circular velocity. When
particles then expand outwards in their epicycle, their tangential velocity reaches a
minimum at their apocenter. The kinematic signature of a resulting positive radial
velocity in the ring will thus be associated with a decreased rotational velocity
towards the outer parts of the ring and a corresponding increased rotational velocity
in the inner parts of the ring. This signature is very significant: the predicted
tangential streaming motions are actually opposite to the sense expected for spiral
arm density waves. Spiral arms cause a tangential velocity increase just outside the
arm and decrease on the inner side of the arm$^{25}$.
\medskip

Streaming motions are indeed observed in M~31 in both the neutral hydrogen gas$^{24}$
and in the CO-traced molecular component$^{2,3}$. The streaming motions are more readily
determined in the diffuse neutral component because the molecular gas is patchier.
Only a moderate spatial resolution is necessary or even desirable because higher-resolution
maps reveal motions on the scale where star formation perturbations (ionized gas expansion,
supernovae driven bubbles, and the like) dominate.
\medskip

In the observed Doppler velocities, the perturbations in the galactocentric radial and
tangential directions are blended in various proportions, depending on the position of
a given point with respect to the major or minor axis. On the major axis only tangential
velocities are measured; on the minor axis, only radial. The inclination of M~31 is large
enough that it can be considered nearly edge-on, but this same high inclination of M~31
makes it difficult to disentangle the radial velocity components of the rings and spiral
arms because both are squeezed in projection along the minor axis. Near the major axis,
though, where tangential perturbations are expected to dominate, projection is less of a
problem. The observed isovelocity curves (see figures 4 and 5) in fact show
wiggles around the 10 kpc ring radius. The sense of these wiggles favours a depression
of tangential velocity in the outer parts of the arm/ring. This favours our interpretation
in terms of a ring induced by collision rather than spiral density waves. In other sectors
of the galaxy, the streaming motions correspond to density-wave features, as expected from
the underlying spiral structure.

\begin{acknowledgements}
This work is based on observations made with the Spitzer Space Telescope,
which is operated by the Jet Propulsion Laboratory, California Institute of
Technology under a contract with NASA. Funding for this work was provided by
by the Anglo American Chairman's Fund as well as by NASA through an award issued by JPL/Caltech.
\bigskip

P.B, G.F., M.A., M.P. and S.W. provided the Spitzer-IRAC data of M31 on which the analyses
are based. D.L.B. initiated the collaboration between the U.S.A,   France and South Africa
and identified Baade dust spirals in emission in the central region of M31.  F.C. discovered
the central ring between the inner Baade arms and believed it to be induced by a head-on
collision. F.B. performed the numerical simulations which fully corroborated this interpretation
and also generated Figures 1 and 2.   R.G. cleaned the IRAC images of foreground stars and
conducted Fourier spectral analysis of the Baade arms. F.C. and F.B. provided crucial drafts
of this Letter. The final revised versions were prepared by D.L.B., S.W, R.G. and M.A.

\end{acknowledgements}

\end{document}